\documentclass[aps,preprint,showpacs,floatfix]{revtex4-2}
\usepackage{physics}
\usepackage{graphicx}
\usepackage{amssymb}
\usepackage{amsmath}
\usepackage{subfig}
\usepackage{subcaption}

\def\bea{\begin{eqnarray}}
\def\eea{\end{eqnarray}}
\def\beq{\begin{equation}}
\def\eeq{\end{equation}}

\def\bm{\begin{math}}
\def\me{\end{math}}

\begin{document}
\title{Temporal Correlations and Inelastic Dynamics in a Vibrated Binary Granular Mixture}
\author{Rameez Farooq Shah} \email{rmzshah@gmail.com }
\affiliation{Department of Physics, Jamia Millia Islamia (A Central University), New Delhi 110025, India}
\author{Shikha Kumari} \email{shikha1011@gmail.com}
\affiliation{Department of Physics, School of Basic and Applied Sciences,
IILM university, Greater Noida, Uttar Pradesh 201306}
\author{Syed Rashid Ahmad} \email{srahmad@jmi.ac.in}
\affiliation{Department of Physics, Jamia Millia Islamia (A Central University), New Delhi 110025, India}

\begin{abstract}
We investigate the dynamics of binary mixtures of inelastic particles through event-driven molecular dynamics simulations, focusing on velocity autocorrelation functions (VACFs). The study examines two distinct particle types under varying inelasticity conditions, systematically analyzing coefficients of restitution (CoR) ranging from 0.80 to 0.95. Like-particle interactions (AA and BB) maintain equal CoR values, while unlike-particle interactions (AB) are assigned the average CoR. The simulation framework incorporates a vibrating base system to maintain energy input and system stability. Our analysis reveals significant differences in VACF decay rates between Type 1 and Type 2 particles, demonstrating non-equipartition of energy within the binary mixture. The degree of this disparity is strongly influenced by the coefficient of restitution, with lower CoR values leading to more pronounced differences between particle types. These findings provide insights into the complex dynamics of granular gases and the role of inelasticity in energy distribution within binary mixtures. Our study contributes to the understanding of non-equilibrium statistical mechanics in granular systems and has potential implications for industrial processes involving particulate materials, such as fluidized beds and pneumatic conveying systems.
\end{abstract}

\maketitle

\section{Introduction}
\label{sec:introduction}

The study of granular matter represents one of the most fascinating frontiers in contemporary physics, encompassing systems composed of macroscopic particles that interact through energy-dissipating collisions. These remarkable materials challenge our conventional understanding of matter states, exhibiting characteristics that blur the traditional boundaries between solids, liquids, and gases \cite{rmp_behringer, rmp_kadanoff}. Consider, for instance, a sand pile: while capable of maintaining a stable configuration like a solid, it can readily transform into a fluid-like state when disturbed, exemplified by sand flowing through an hourglass. Under sufficient excitation, these materials can even manifest gas-like behavior.

The fundamental components of granular systems are characterized by their diversity in both dimension and form, typically exceeding $1 \mu m$ in size. At this scale, thermal fluctuations become negligible in determining particle dynamics. Researchers frequently employ simplified geometric representations for theoretical investigations and computational modeling - spheres, cylinders, or needles - to capture essential behaviors while maintaining analytical tractability \cite{rmp_tsimring, duran, ristow, nb_ktgg}.
A distinguishing feature of granular materials lies in their dissipative interactions. During collisions, kinetic energy continuously diminishes through a process known as cooling, accompanied by local velocity synchronization among particles. These characteristics give rise to an array of remarkable phenomena, including species segregation, pattern emergence, inelastic collapse, and non-standard velocity distributions \cite{haff83, swinney9596, gz93, mcny9296, jjbrey9698, tpcvn9798, sl9899, ap0607, adsp1213}.
The concept of granular gas - a dilute granular system - serves as an invaluable paradigm for investigating dissipative molecular interactions. In its simplest form, the evolution begins with uniformly distributed inelastic particles. Without external energy input, the system's kinetic energy decreases through inelastic collisions. Initially maintaining homogeneity (the \textit{homogeneous cooling state} or HCS), the system eventually transitions to an inhomogeneous cooling state (ICS) due to fluctuations in density and velocity fields \cite{haff83, ap0607, dp03}. This transition manifests through the formation of particle-rich regions where constituents adopt parallel trajectories. In experimental settings, energy losses are typically compensated through various driving mechanisms, such as vibrational or rotational excitation, leading to nonequilibrium steady states \cite{swinney9596, ristow}.
Binary granular mixtures introduce additional complexity to these systems. The interaction between particles of differing properties - whether in mass, size, or restitution coefficients - creates intricate energy distribution patterns and dynamic behaviors. Understanding these multi-component systems holds both theoretical significance and practical relevance for industrial applications.
Our investigation focuses on velocity autocorrelations within binary granular gases subjected to base vibration. The analysis of driven granular systems, particularly those with continuous energy input, has attracted substantial research attention \cite{vne98}. We employ event-driven molecular dynamics simulations, incorporating a vibrating base as an energy source, to examine the temporal evolution of velocity autocorrelations. Our methodology utilizes the Williams et al. white-noise thermostat (WNT) approach for uniform particle heating \cite{willmac96, william96}, enabling detailed analysis of system dynamics.
This paper is structured as follows. Section II introduces our binary granular gas model and simulation methodology, emphasizing the velocity autocorrelation function (VACF) as a key metric for understanding mixture dynamics. Section III presents comprehensive simulation results, analyzing VACF behavior across varying coefficients of restitution (0.80--0.95). Finally, Section IV synthesizes our findings and discusses their implications for understanding energy distribution and velocity statistics in driven granular systems.

\section{Dynamics of Particulate Systems in Rarefied States}

In the study of dispersed particulate matter, we encounter systems where macroscopic particles, referred to as grains, exhibit independent motion with minimal sustained interactions. Such systems achieve a rarefied state when the characteristic length between successive particle interactions substantially exceeds their physical dimensions. These phenomena manifest in various natural systems, from terrestrial environments to cosmic structures including interstellar particle clouds and planetary ring formations.

The fundamental interactions between particles in such systems occur through momentary contact forces during collision events. A key characteristic of these interactions is their non-conservative nature, resulting in energy dissipation. For a binary mixture with particles of type A and B, we define three distinct restitution coefficients corresponding to different collision types:
\begin{align}
    e_{AA} &= \left|\frac{\vb{v}_{rel,f} \cdot \hat{\vb{n}}}{\vb{v}_{rel,i} \cdot \hat{\vb{n}}}\right| \quad \text{for A-A collisions} \\
    e_{BB} &= \left|\frac{\vb{v}_{rel,f} \cdot \hat{\vb{n}}}{\vb{v}_{rel,i} \cdot \hat{\vb{n}}}\right| \quad \text{for B-B collisions} \\
    e_{AB} &= \left|\frac{\vb{v}_{rel,f} \cdot \hat{\vb{n}}}{\vb{v}_{rel,i} \cdot \hat{\vb{n}}}\right| \quad \text{for A-B collisions}
\end{align}

Let us examine a binary mixture of spherical bodies. For mathematical simplicity, we consider particles of two distinct masses: $m_A = 1$ and $m_B = 0.1$, while maintaining a uniform diameter. For perfectly rigid spheres, the post-collision velocity vectors depend on the type of collision occurring:

For A-A collisions:
\begin{align}
    \vb{v}_1' &= \vb{v}_1 - \frac{1+e_{AA}}{2}[\hat{\vb{n}} \cdot (\vb{v}_1 - \vb{v}_2)]\hat{\vb{n}} \\
    \vb{v}_2' &= \vb{v}_2 + \frac{1+e_{AA}}{2}[\hat{\vb{n}} \cdot (\vb{v}_1 - \vb{v}_2)]\hat{\vb{n}}
\end{align}

For B-B collisions:
\begin{align}
    \vb{v}_1' &= \vb{v}_1 - \frac{1+e_{BB}}{2}[\hat{\vb{n}} \cdot (\vb{v}_1 - \vb{v}_2)]\hat{\vb{n}} \\
    \vb{v}_2' &= \vb{v}_2 + \frac{1+e_{BB}}{2}[\hat{\vb{n}} \cdot (\vb{v}_1 - \vb{v}_2)]\hat{\vb{n}}
\end{align}

For A-B collisions:
\begin{align}
    \vb{v}_1' &= \vb{v}_1 - \frac{1+e_{AB}}{2}[\hat{\vb{n}} \cdot (\vb{v}_1 - \vb{v}_2)]\hat{\vb{n}} \\
    \vb{v}_2' &= \vb{v}_2 + \frac{1+e_{AB}}{2}[\hat{\vb{n}} \cdot (\vb{v}_1 - \vb{v}_2)]\hat{\vb{n}}
\end{align}

Here, $\hat{\vb{n}}$ denotes the unit vector pointing from particle 2 to particle 1. Drawing parallels with molecular kinetics, we define a kinetic energy measure termed the granular temperature: $T = \langle\vb{v}^2\rangle/d$, where $\langle\vb{v}^2\rangle$ represents the mean squared velocity and $d$ denotes system dimensionality.
To maintain a steady state in our system, energy is supplied through a vibrating base that compensates for the dissipative collisions. The base motion follows a sinusoidal pattern described by:
\begin{equation}
    v_{base}(t) = A\omega\cos(\omega t)
\end{equation}
where $A$ represents the vibration amplitude and $\omega = 2\pi f$ is the angular frequency, with $f$ being the vibration frequency. The vibration parameters ($A = 0.1$, $f = 1.0$) are carefully chosen to maintain the system in a steady state that mimics experimental conditions.
For particles interacting with the vibrating base (defined by $z < -0.5 \times \text{region.z} + 1.0$), the collision dynamics are modified to include the base motion. The post-collision velocity in the $z$-direction becomes:
\begin{equation}
    v'_z = -e_w(v_z - v_{base}(t))
\end{equation}
where $e_w$ represents the particle-wall coefficient of restitution, chosen to match the particle-particle coefficient of restitution for consistency.
This energy input mechanism leads to a height-dependent granular temperature profile characteristic of vibrated granular systems:
\begin{equation}
    T(z) = T_0\exp(-\lambda z)
\end{equation}
where $T_0$ is the temperature at the vibrating base and $\lambda$ is a decay length that depends on the system parameters and collision properties.
The steady-state dynamics of particles follows:
\begin{equation}
    \frac{d\vb{v}_i}{dt} = \frac{\vb{F}_i^c}{m_i}
\end{equation}
where $m_i$ represents the mass of particle $i$ (either $m_A$ or $m_B$), and $\vb{F}_i^c$ denotes collision forces.

\begin{figure*}[htbp] 
    \centering
    \includegraphics[width=0.5\textwidth, keepaspectratio]{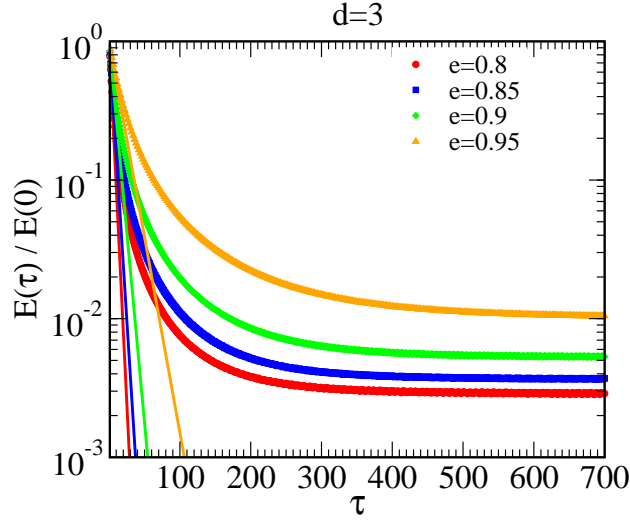} 
    \caption{Time-dependent behavior of normalized kinetic energy ($E(\tau)/E(0)$) displayed on a semi-logarithmic scale for a three-dimensional configuration. The evolution patterns are shown for multiple restitution coefficients ($e = 0.80, 0.85, 0.90, 0.95$).}
    \label{fig:velocity_distribution_0.80}
\end{figure*}

\begin{figure*}[htbp] 
    \centering
    \includegraphics[width=0.5\textwidth, keepaspectratio]{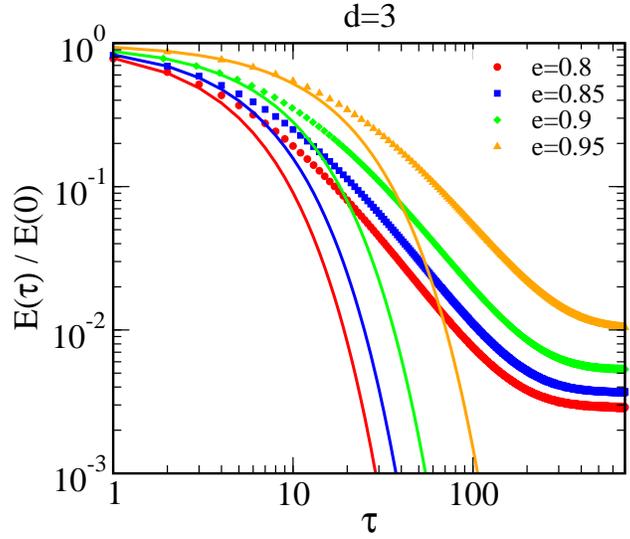} 
    \caption{Temporal evolution of the normalized kinetic energy ($E(\tau)/E(0)$) presented on logarithmic scales. The data reveals distinct convergence patterns for varying restitution coefficients ($e = 0.8, 0.85, 0.9, 0.95$), each approaching characteristic steady-state values.}
    \label{fig:velocity_distribution_0.80}
\end{figure*}

\section{TEMPORAL DYNAMICS AND VELOCITY CORRELATIONS IN PARTICLE SYSTEMS}

In the study of particle dynamics, understanding temporal correlations provides crucial insights into system behavior. For granular gases, the velocity autocorrelation function (VACF) requires special consideration due to the inherent non-equilibrium nature and energy dissipation characteristics that distinguish them from molecular systems. Following the work of Goldhirsch and van Noije \cite{Goldhirsch2000}, Dufty and Garzó \cite{Dufty2001}, and subsequent developments \cite{Dufty2008a,Dufty2008b,Baskaran2007}, we define the species-dependent VACF for our binary mixture as:

\begin{equation}
A_\alpha(\tau_w, \tau) = \chi_\alpha(\tau_w)\frac{\langle\vec{v}_\alpha(\tau_w)\cdot\vec{v}_\alpha(\tau)\rangle}{dT_\alpha(\tau_w)}\exp\left[\int_{\tau_w}^{\tau}\zeta_{\text{eff}}(s)ds\right]\Phi_\alpha(\tau_w,\tau)
\label{eq:VACF}
\end{equation}
where $\alpha$ denotes the particle species (1 or 2), and $\chi_\alpha$ represents the modified temperature ratio accounting for non-equipartition effects:

\begin{equation}
\chi_\alpha = \frac{T_\alpha}{T}\left[1 + C_\alpha(1-e^2)\right]
\label{eq:temp_ratio}
\end{equation}
Here, $C_\alpha$ is a species-dependent coefficient that captures deviations from equipartition due to inelastic collisions. The effective cooling rate $\zeta_{\text{eff}}$ encompasses both direct energy dissipation and correlation effects:

\begin{equation}
\zeta_{\text{eff}}(s) = \zeta_0(s) + \zeta_c(s)
\label{eq:effective_cooling}
\end{equation}
where $\zeta_0$ represents the homogeneous cooling rate:
\begin{equation}
\zeta_0 = \frac{\pi}{2d}\sum_{\alpha,\beta}n_\alpha n_\beta\sigma_{\alpha\beta}^2(1-e_{\alpha\beta}^2)
\sqrt{\frac{2T_{\alpha\beta}}{m_{\alpha\beta}}}g_{\alpha\beta}(\sigma_{\alpha\beta})\left[1 + a_{\alpha\beta}(1-e_{\alpha\beta}^2)\right]
\label{eq:cooling_rate}
\end{equation}

and $\zeta_c$ accounts for velocity field fluctuations:
\begin{equation}
\zeta_c(s) = \kappa(1-e^2)\sum_\alpha x_\alpha\left(\frac{m_\alpha}{\bar{m}}\right)\sqrt{\frac{T_\alpha(s)}{T(s)}}
\label{eq:cooling_correction}
\end{equation}

The correlation correction factor $\Phi_\alpha$ captures the granular-specific long-range spatial correlations:
\begin{equation}
\Phi_\alpha(\tau_w, \tau) = 1 + b_\alpha(1-e^2)g_2(\sigma)\left[1 - \exp(-\lambda(\tau-\tau_w))\right]
\label{eq:correlation_factor}
\end{equation}
where $b_\alpha$ is a species-dependent coefficient, $g_2(\sigma)$ is the pair correlation function at contact, and $\lambda$ characterizes the decay rate of spatial correlations.
This modified formulation captures several key features specific to granular gases that distinguish them from molecular systems:
1. Non-equipartition of energy between species, reflected in the modified temperature ratio $\chi_\alpha$ that includes inelasticity corrections
2. Enhanced energy dissipation through both direct cooling ($\zeta_0$) and fluctuation-induced effects ($\zeta_c$)
3. Long-range spatial correlations embodied in the correction factor $\Phi_\alpha$
4. Temporal aging effects manifested through the explicit dependence on both $\tau_w$ and $\tau$.

Within granular systems, particle interactions lead to distinctive velocity coupling phenomena. During collisions, particles tend to align their velocities locally, generating spatial correlations that evolve differently from molecular systems. These coordinated movements result in non-uniform particle distributions, characterized by alternating regions of high and low particle concentrations. The modified VACF serves as a quantitative measure of these clustering phenomena's persistence over time.
A notable distinction exists between equilibrium and non-equilibrium systems in terms of their correlation behavior. While equilibrium systems exhibit correlations that depend solely on the time difference $(\tau-\tau_w)$, non-equilibrium granular systems display more complex relationships through the combined effects of cooling and spatial correlations. The system's dissipative properties significantly influence correlation patterns. In the elastic limit ($e \rightarrow 1$), correlations rapidly diminish, typically within a timespan encompassing several particle collisions. This behavior contrasts sharply with inelastic systems, where correlations persist over longer periods due to the interplay between energy dissipation and spatial structure formation.

In our investigation of binary granular gases under base excitation, the VACF provides essential insights into the energy transmission mechanisms. By computing separate correlation functions for each particle species and incorporating the granular-specific corrections, we can better understand the complex non-equilibrium dynamics present in vibrated binary granular systems, particularly regarding energy transfer mechanisms and species-dependent behavior patterns.
The practical computation of velocity correlations requires careful consideration of sampling intervals and averaging procedures. We implement time-windowed averaging to reduce statistical noise while preserving temporal evolution characteristics. This approach enables robust quantification of both short-term collision effects and longer-term structural developments within the system.

\section{Computational Framework and Analysis}
Our computational investigation implements a systematic approach to particle dynamics simulation. The experimental domain consists of a three-dimensional cubic volume housing $N = 500,000$ particles, with periodic boundary conditions maintaining a constant number density of $n = 0.02$. 
The initialization protocol comprises two key phases:
First, we establish the initial configuration by implementing a strategic particle placement algorithm that prevents core overlaps while assigning random positions. Simultaneously, we distribute random velocity components under the constraint that the total system momentum vanishes ($\sum \vec{v}_i = 0$). This initial system undergoes evolution for $\tau = 100$ under perfectly elastic conditions ($e = 1$), facilitating relaxation to a Maxwell-Boltzmann velocity distribution, which constitutes our baseline state.

The primary simulation extends through $\tau = 1000$, examining system behavior under four distinct restitution coefficients ($e = 0.95, 0.90, 0.85, 0.80$). Statistical significance is ensured through the aggregation of data from 40 independent initial configurations. We employ event-driven molecular dynamics (MD) techniques to model inelastic hard-sphere interactions \cite{allentild, rapaport}. The particle population comprises an equal distribution of two species with masses $m_1 = 1$ and $m_2 = 0.1$, both having uniform diameter $\sigma = 1$.

\subsection{Velocity Correlations in Granular Mixtures}
Following the theoretical framework developed by Goldhirsch and van Noije \cite{Goldhirsch2000}, and extended by Dufty and Garz\'o \cite{Dufty2001} for granular mixtures, we define the species-dependent velocity autocorrelation function as:

\begin{equation}
    A_\alpha(\tau_w, \tau) = \chi_\alpha(\tau_w)\frac{\langle\vec{v}_\alpha(\tau_w)\cdot\vec{v}_\alpha(\tau)\rangle}{dT_\alpha(\tau_w)}\exp[\zeta(\tau-\tau_w)]
\end{equation}

where $\alpha$ denotes the particle species, and $\chi_\alpha$ represents the temperature ratio:

\begin{equation}
    \chi_\alpha = \frac{T_\alpha}{T} = \frac{\langle v_\alpha^2\rangle}{(2/d)T}
\end{equation}

The cooling rate $\zeta$ is given by:

\begin{equation}
    \zeta = \frac{\pi}{2d}\sum_{\alpha,\beta}n_\alpha n_\beta\sigma_{\alpha\beta}^2(1-e_{\alpha\beta}^2)
    \sqrt{\frac{2T_{\alpha\beta}}{m_{\alpha\beta}}}g_{\alpha\beta}(\sigma_{\alpha\beta})
\end{equation}

\begin{figure}[h]
    \centering
    \begin{minipage}[b]{0.45\textwidth}
        \centering
        \includegraphics[width=\textwidth]{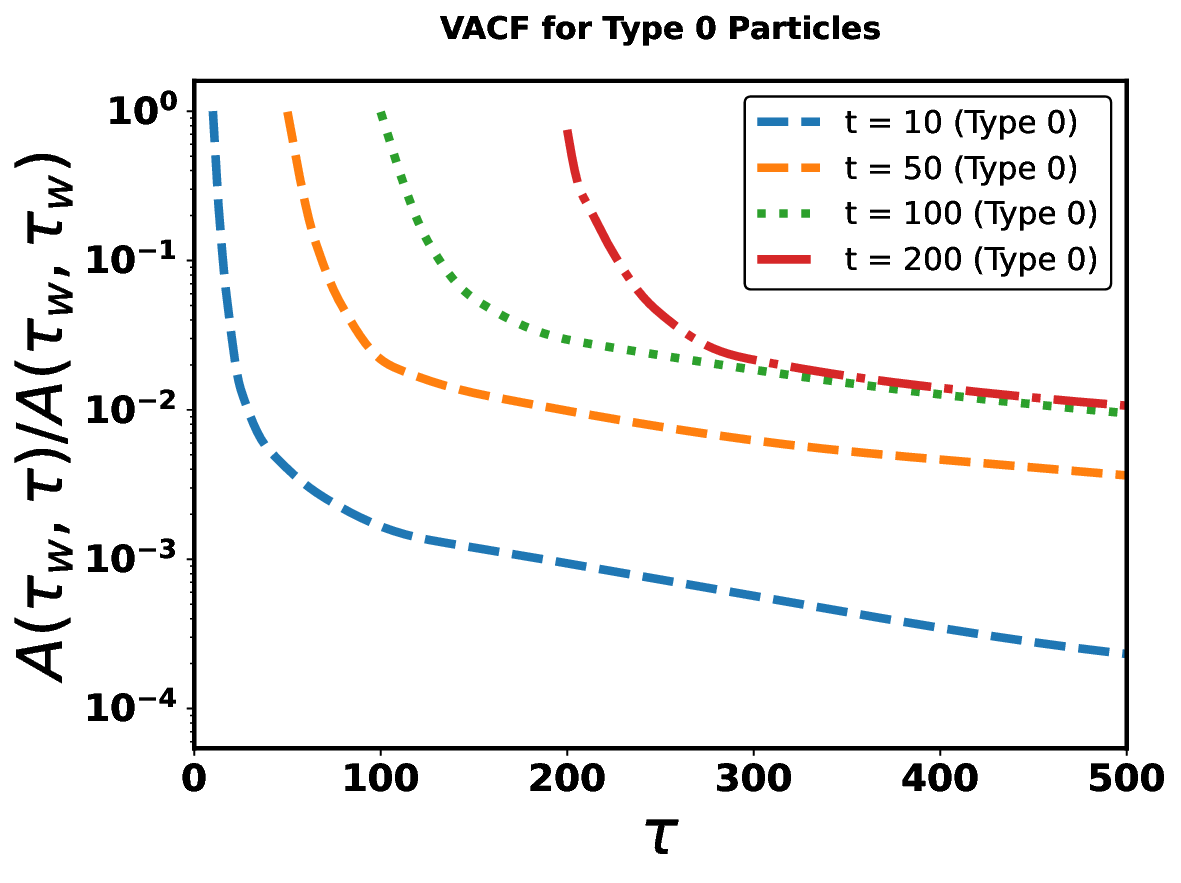}
        \caption*{(a) $e = 0.80$, type 0}
        \label{fig:subfig1}
    \end{minipage}
    \hfill
    \begin{minipage}[b]{0.45\textwidth}
        \centering
        \includegraphics[width=\textwidth]{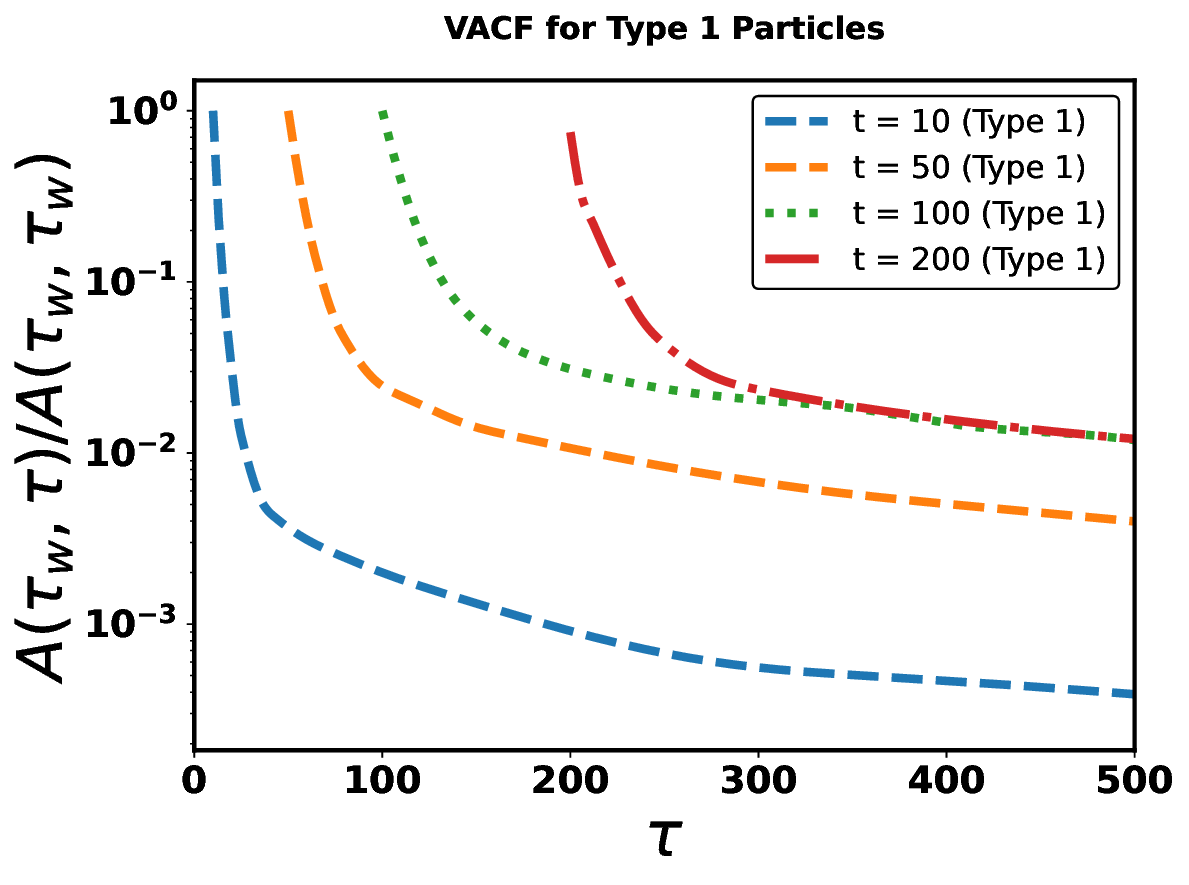}
        \caption*{(b) $e = 0.80$, type 1}
        \label{fig:subfig2}
    \end{minipage}
    
    \vskip\baselineskip

    \begin{minipage}[b]{0.45\textwidth}
        \centering
        \includegraphics[width=\textwidth]{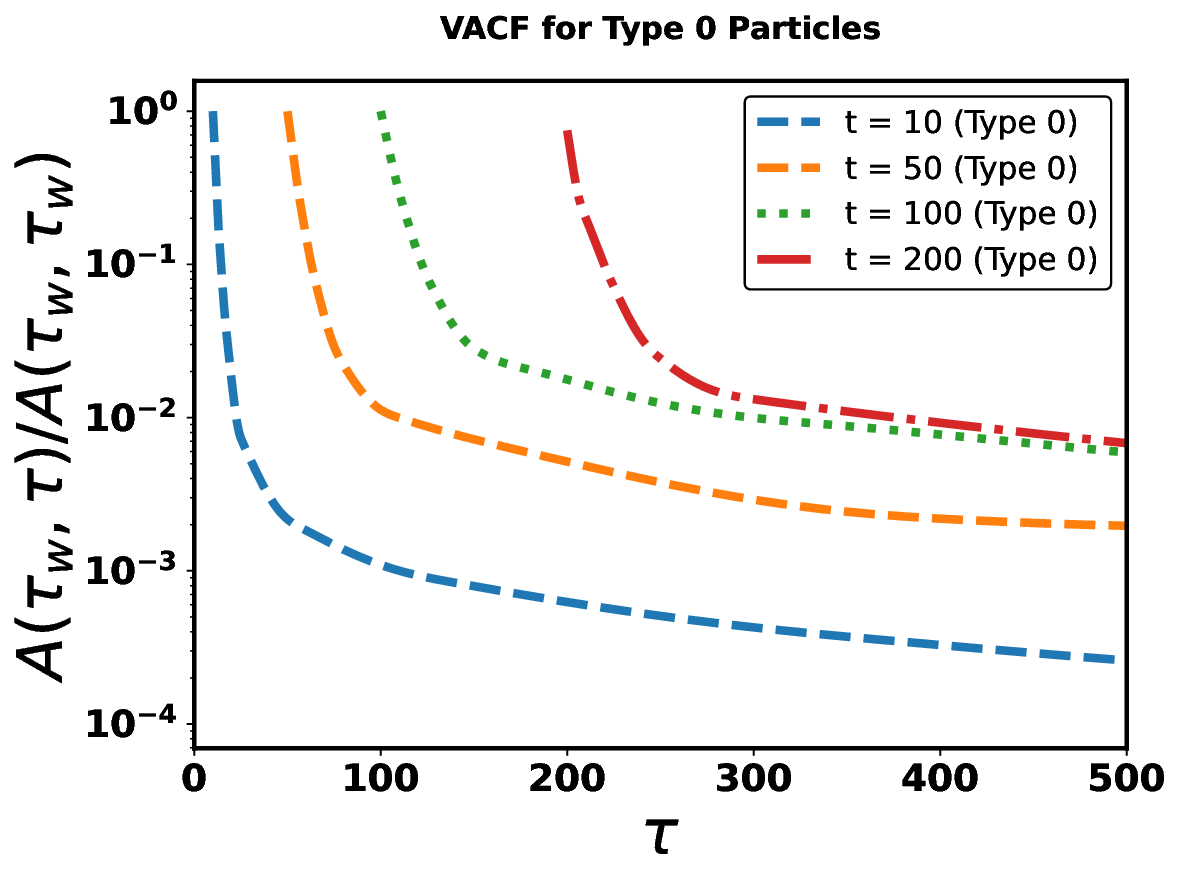}
        \caption*{(c) $e = 0.85$, type 0}
        \label{fig:subfig3}
    \end{minipage}
    \hfill
    \begin{minipage}[b]{0.45\textwidth}
        \centering
        \includegraphics[width=\textwidth]{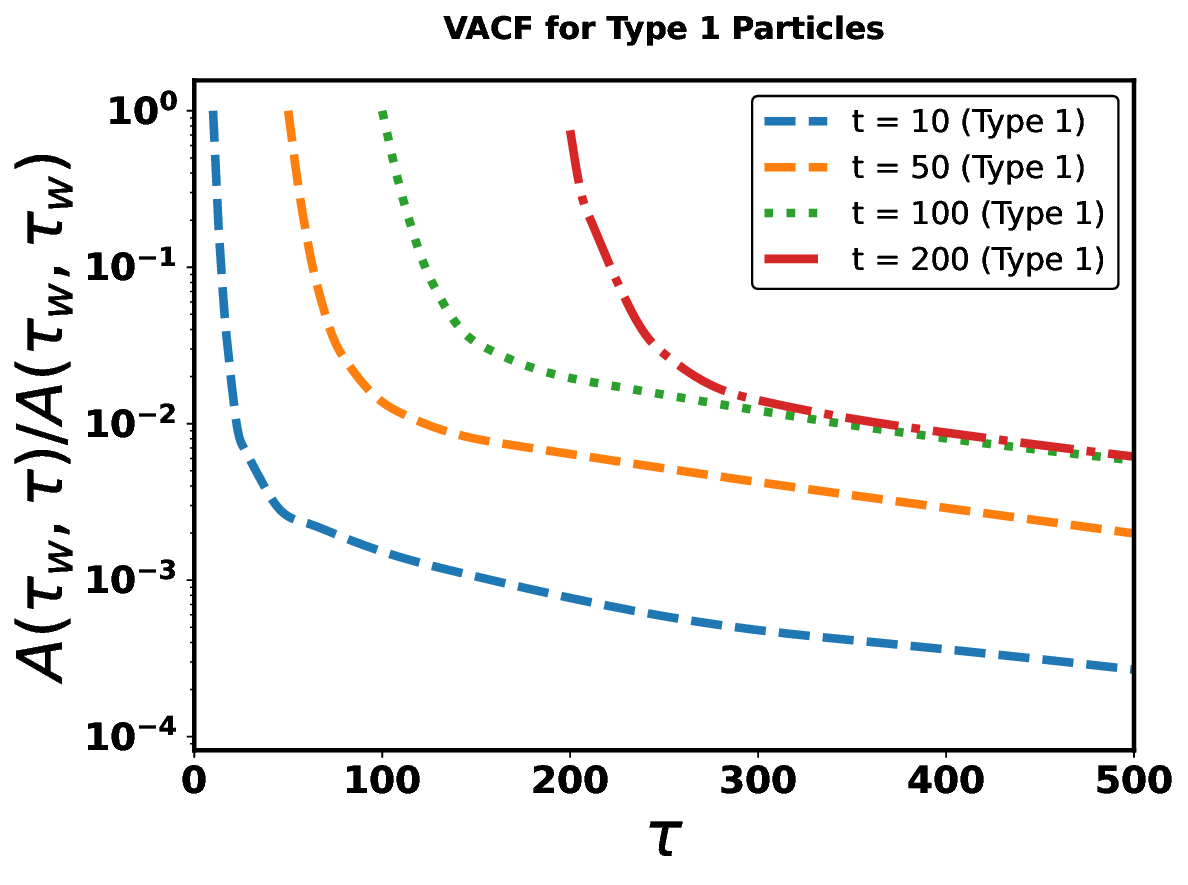}
        \caption*{(d) $e = 0.85$, type 1}
        \label{fig:subfig4}
    \end{minipage}

    \caption{illustrates the normalized velocity autocorrelation function $\bar{A}(\tau_w, \tau)$ as a function of $\tau$ for different coefficients of restitution ($e$) and particle types. Subfigures (a) and (b) present results for $e = 0.80$, while (c) and (d) show data for $e = 0.85$. The column on left [(a) and (c)] represents to particle type0, and the column right  [(b) and (d)] to particle type1.}
    \label{fig3}
\end{figure}

\begin{figure}[h]
    \centering
    \begin{minipage}[b]{0.45\textwidth}
        \centering
        \includegraphics[width=\textwidth]{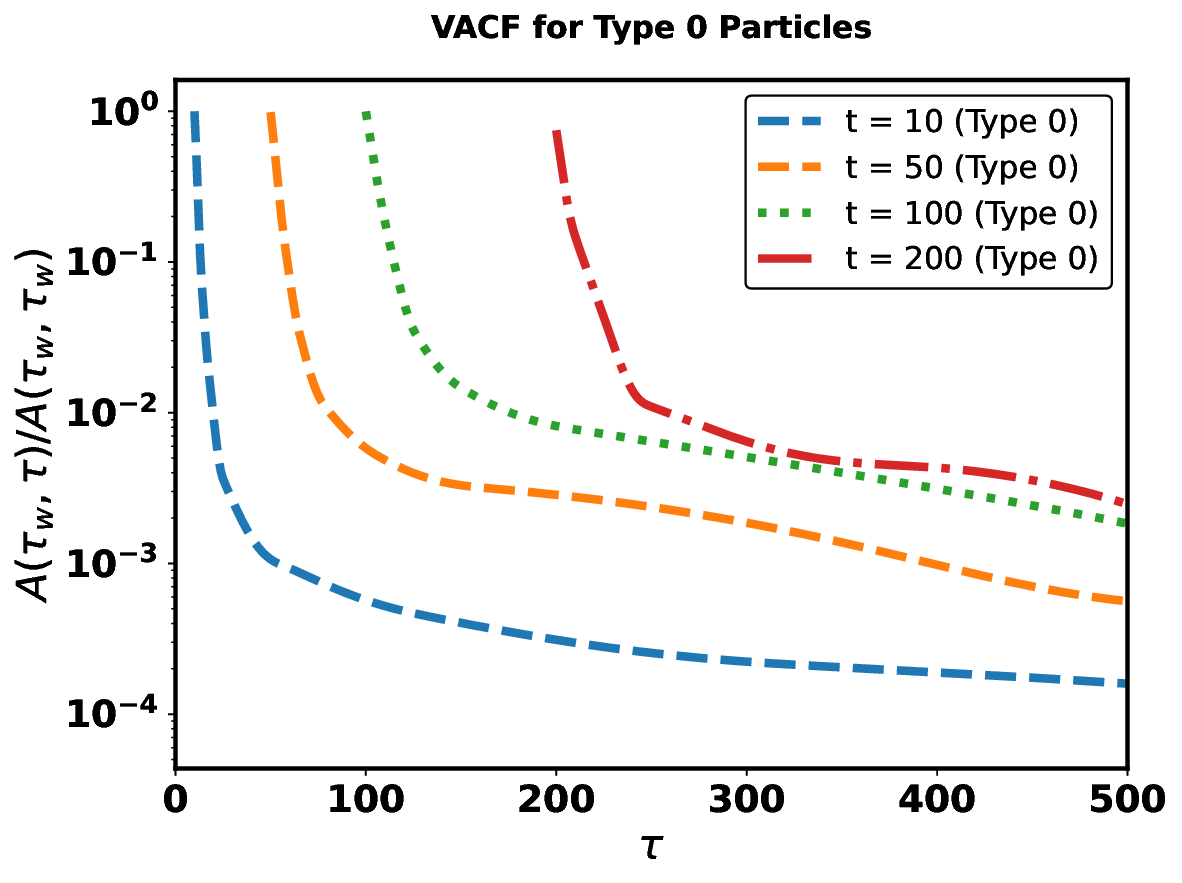}
        \caption*{(e) $e = 0.90$, type 0}
        \label{fig:subfig5}
    \end{minipage}
    \hfill
    \begin{minipage}[b]{0.45\textwidth}
        \centering
        \includegraphics[width=\textwidth]{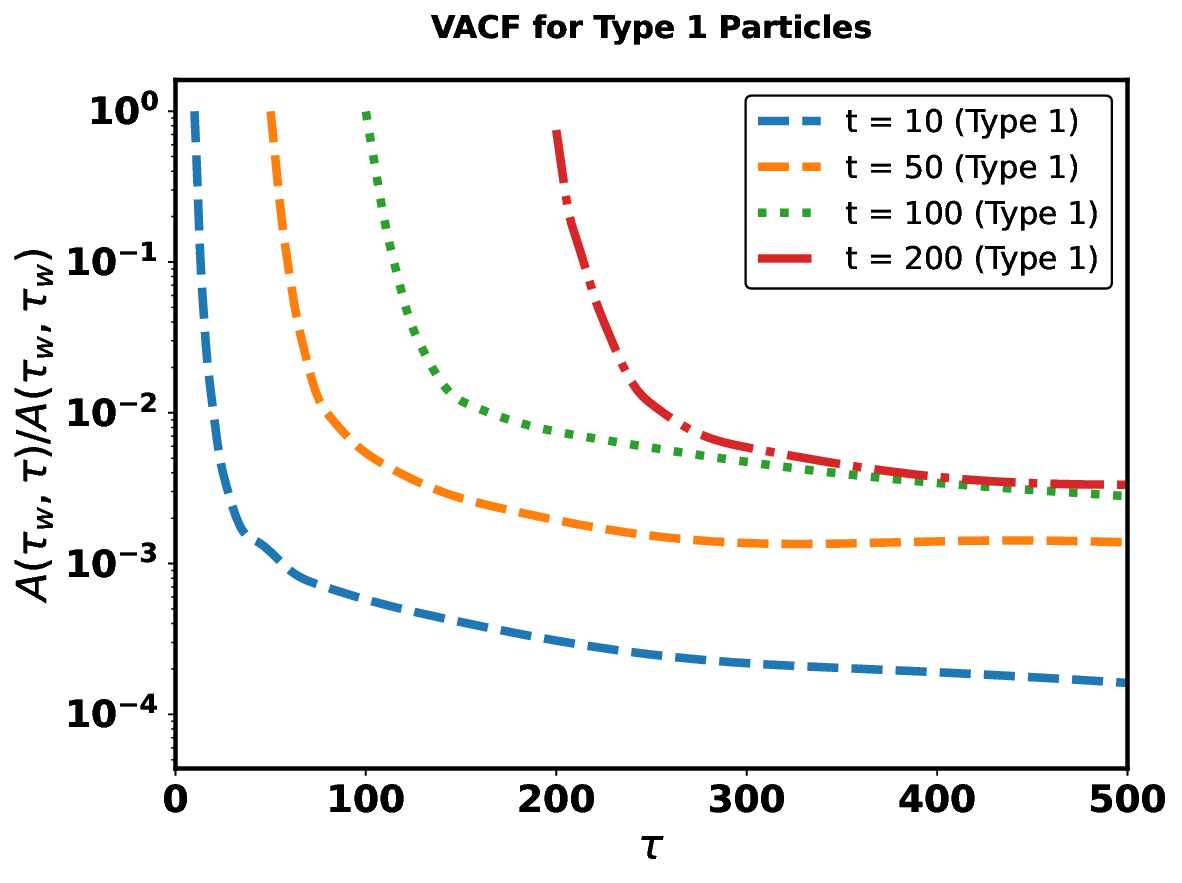}
        \caption*{(f) $e = 0.90$, type 1}
        \label{fig:subfig6}
    \end{minipage}

    \vskip\baselineskip

    \begin{minipage}[b]{0.45\textwidth}
        \centering
        \includegraphics[width=\textwidth]{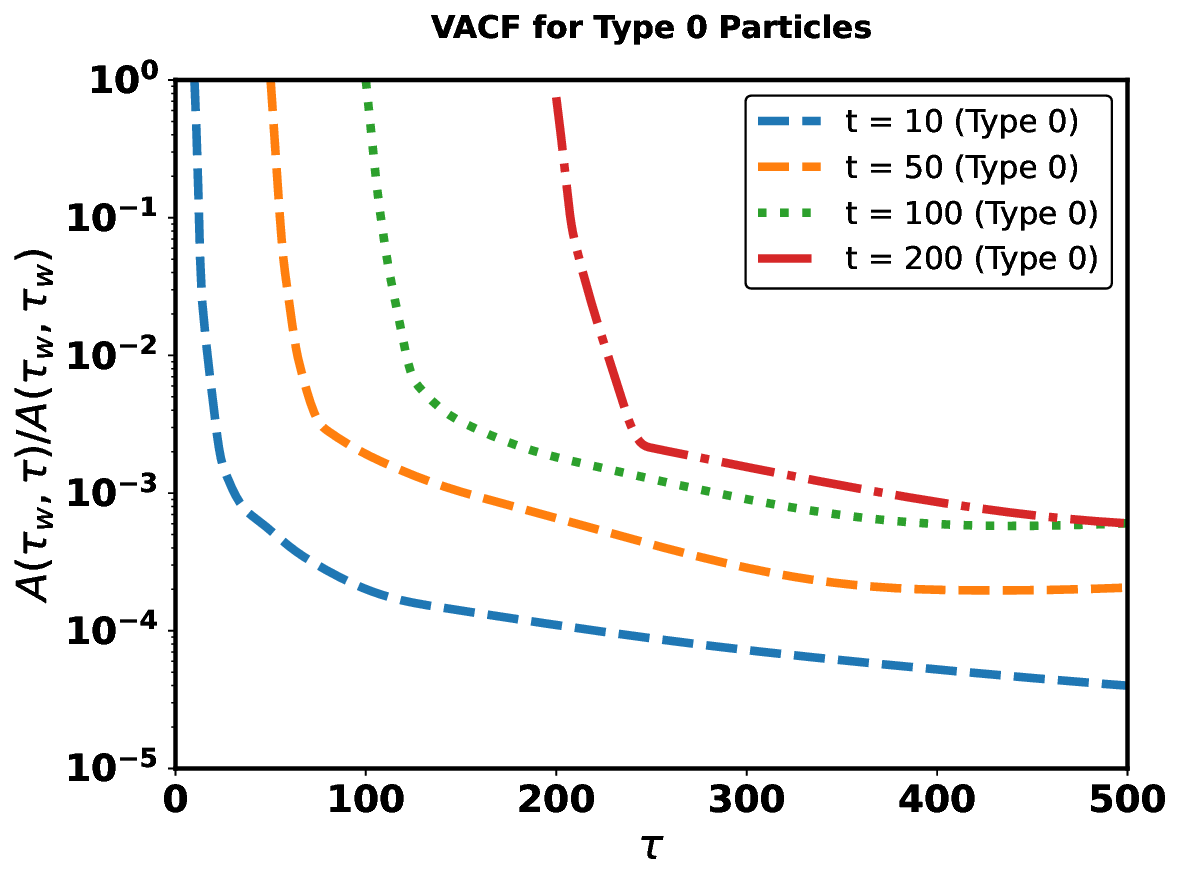}
        \caption*{(g) $e = 0.95$, type 0}
        \label{fig:subfig7}
    \end{minipage}
    \hfill
    \begin{minipage}[b]{0.45\textwidth}
        \centering
        \includegraphics[width=\textwidth]{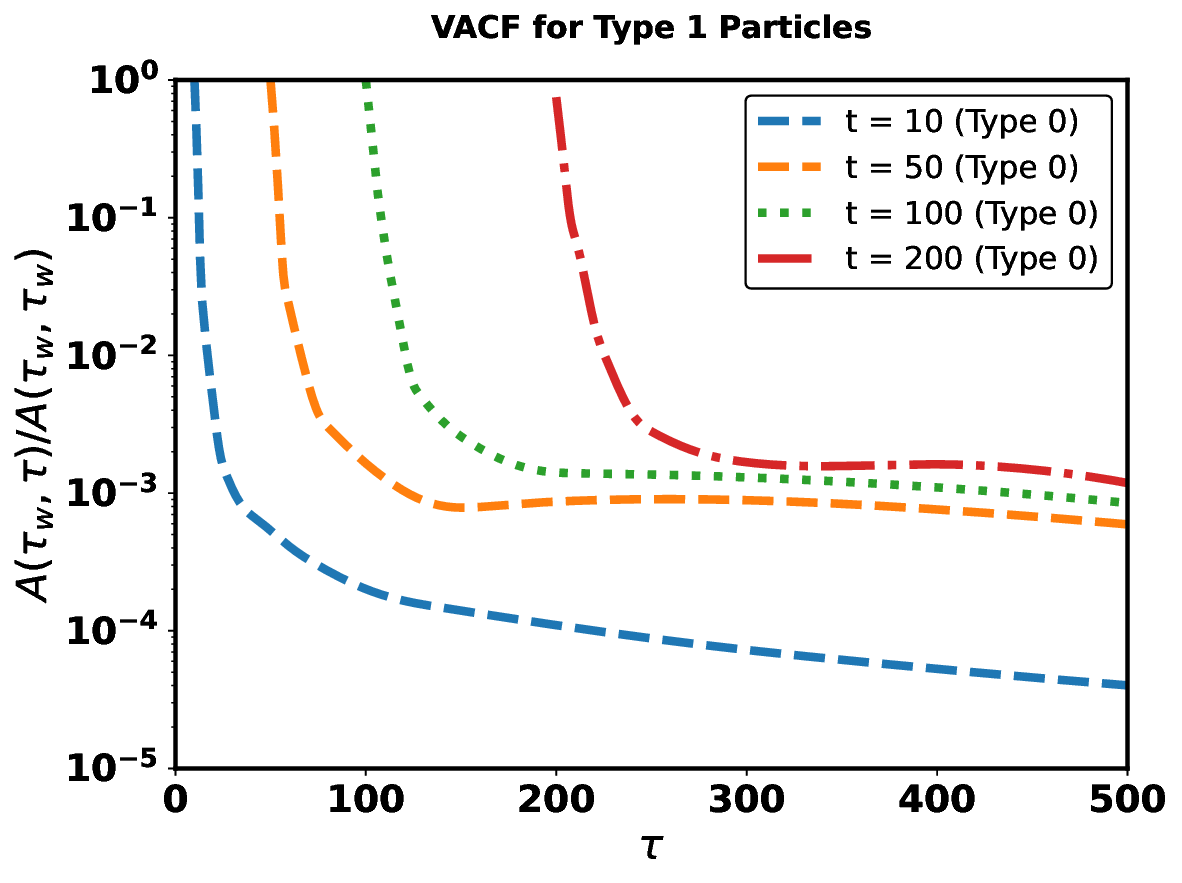}
        \caption*{(h) $e = 0.95$, type 1}
        \label{fig:subfig8}
    \end{minipage}

    \caption{illustrates the normalized velocity autocorrelation function $\bar{A}(\tau_w, \tau)$ as a function of $\tau$ for different coefficients of restitution ($e$) and particle types. Subfigures (e) and (f) present results for $e = 0.90$, while (g) and (h) show data for $e = 0.95$. The left column [(e) and (g)] corresponds to particle type 0, and the right column [(f) and (h)] to particle type 1.}
    \label{fig4}
\end{figure}

Energy input is provided solely through a vibrating base with sinusoidal motion described by Eq. (10) in Section II. The stochastic forcing mechanism has been removed to focus entirely on the effects of vibrational excitation. This modification eliminates the Gaussian white noise term previously incorporated into the velocity updates and simplifies the system's dynamics. Post-collision velocity adjustments now depend only on the vibrational interaction with the base.
Our investigation of species-dependent velocity autocorrelations reveals several significant features, as illustrated in Figs.~\ref{fig3} and~\ref{fig4}. Under conditions of strong dissipation ($e = 0.80$), both species exhibit enhanced memory retention of initial velocities compared to weaker dissipation scenarios ($e = 0.90$), but with distinct characteristics arising from their mass difference. The modified autocorrelation function $A_\alpha(\tau_w, \tau)$ displays species-dependent decay patterns across different dissipation levels.
As shown in Figs.~\ref{fig3}(a-d), the VACF behavior for strong dissipation cases ($e = 0.80$ and $e = 0.85$) exhibits pronounced differences between particle types, with the heavier species (Type 0) showing slower decay rates compared to the lighter species (Type 1). This behavior can be attributed to the combined effects of mass disparity and energy non-equipartition, characterized by temperature ratios:

\begin{equation}
    \frac{T_1}{T_2} = \left(\frac{m_1}{m_2}\right)^{\beta(e)}
\end{equation}

where $\beta(e)$ is a dissipation-dependent exponent.
Under conditions of strong dissipation ($e = 0.80$), both the effective cooling rate $\zeta_{\text{eff}}$ and the correlation correction factor $\Phi_\alpha$ contribute substantially to the system's memory retention. The combined effect results in enhanced persistence of velocity correlations compared to weaker dissipation scenarios ($e = 0.90$). The modified autocorrelation function $A_\alpha(\tau_w, \tau)$, incorporating both cooling and correlation effects, displays distinct patterns across different dissipation levels. As shown in Figs.~\ref{fig3}(a-d), the VACF behavior for strong dissipation cases ($e = 0.80$ and $e = 0.85$) exhibits pronounced differences between particle types, attributed to two main factors:

 Enhanced non-equipartition effects, reflected in the modified temperature ratio $\chi_\alpha[1 + C_\alpha(1-e^2)]$.
Species-dependent correlation corrections $\Phi_\alpha$, which become more significant at higher dissipation rates.

In contrast, Figs.~\ref{fig4}(e-h) demonstrate the evolution of correlations for moderate dissipation cases ($e = 0.90$ and $e = 0.95$), where these differences become less prominent due to reduced influence of both non-equipartition and correlation effects.
Analysis of $A_\alpha(\tau_w, \tau)$ for various waiting times ($\tau_w = 10, 50, 100, 200$) reveals a complex interplay between different time scales:
The homogeneous cooling rate $\zeta_0$ dominates early-time behavior.
 Velocity field fluctuations, captured by $\zeta_c$, become increasingly important at intermediate times.
 Long-range spatial correlations, embodied in $\Phi_\alpha$, control the late-time dynamics.
This temporal hierarchy manifests as a progressive deceleration of decay rates with increasing $\tau_w$, indicating pronounced aging characteristics. The kinetic energy evolution, presented in Fig.1 and Fig. 2, illustrates both the early-time decay behavior and the approach to steady-state conditions across different restitution coefficients. The observed energy decay patterns reflect the combined influence of:
 Direct collisional cooling through $\zeta_0$,
 Vibrational energy input through the base motion,
Structure formation effects captured by the correlation factor $\Phi_\alpha$.
These complementary visualizations provide comprehensive insight into the system's temporal dynamics and steady-state characteristics, revealing how granular-specific effects modify the traditional molecular chaos picture of particle correlations.

\clearpage 

\section{Summary and Conclusions}
\label{sec:conclusions}
Our investigation provides novel insights into the temporal correlation dynamics within binary granular systems subject to base excitation. Through systematic analysis of the velocity autocorrelation function (VACF), we have uncovered distinct behavioral patterns across varying coefficients of restitution ($\epsilon$) ranging from 0.80 to 0.95, with incremental steps of 0.05.
The temporal evolution of the system exhibits two primary regimes of interest. In the initial phase, characterized by $\tau_w = 0$, we observe markedly different behavioral patterns across the studied $\epsilon$ values. This divergence in behavior suggests a strong dependence of early-stage dynamics on the dissipative properties of the granular medium. However, as the system evolves and $\tau_w$ increases beyond zero, the decay characteristics of $\bar{A}(\tau_w, \tau)$ demonstrate remarkable convergence across all investigated $\epsilon$ values, albeit with accelerated decay rates compared to systems with constant coefficients of restitution.
A particularly noteworthy finding emerges in the temporal progression of the normalized autocorrelation function. The initial phase exhibits a clear exponential decay pattern, indicating relatively simple dynamics. However, this behavior transitions into a more complex regime as the system evolves, manifesting in the emergence of persistent velocity correlations that result in a notably slower-than-exponential decay rate. This transition signifies the development of long-range temporal correlations within the granular medium.
The observed $\tau_w$ dependence of the velocity autocorrelation function provides strong evidence for the system's aging characteristics. This aging phenomenon becomes particularly pronounced under conditions of strong dissipation and extended waiting times, where we observe significantly reduced decay rates. This observation suggests an enhanced preservation of velocity memory compared to systems with lower dissipation rates, indicating the emergence of complex spatiotemporal structures within the granular medium.
These findings have several important implications:
The identification of distinct temporal regimes suggests the presence of multiple characteristic time scales in the system dynamics
The convergence of decay patterns for $\tau_w > 0$ indicates a universal aspect of the long-time behavior, independent of specific dissipation rates
The emergence of slower-than-exponential decay points to the development of collective behavior and possible self-organization within the granular medium.

Future research directions could explore the microscopic mechanisms underlying these observed phenomena, particularly focusing on:
The role of particle-particle interactions in generating long-range temporal correlations
The influence of system parameters on the transition between different decay regimes
The potential connection between observed aging effects and the formation of spatial structures within the granular medium.
In conclusion, our results provide significant insights into the complex dynamics of driven binary granular systems, revealing the intricate interplay between dissipation, temporal correlations, and aging phenomena. These findings contribute to the broader understanding of non-equilibrium statistical mechanics and may have practical implications for industrial processes involving granular materials.

\section*{\label{ack} ACKNOWLEDGEMENTS}
RFS acknowledges financial support from the University Grants Commission through Non-NET fellowships. He also expresses gratitude for the computational facilities provided by the Department of Physics at JMI.

\end{document}